%% file: main.tex
\documentclass[conference]{IEEEtran}
\IEEEoverridecommandlockouts

\usepackage{graphicx} %
\usepackage{url}
\usepackage{color}
\usepackage[capitalise,noabbrev]{cleveref} 
\crefname{appsec}{Appendix}{Appendices} 
\usepackage{subcaption}

\newcommand{\name}{ScooterLab}

\title{\name: A Programmable and Participatory Sensing Research Testbed using \\Micromobility Vehicles

\thanks{This work is supported by National Science Foundation (NSF) under award number 2234516.}
}

\author{
\IEEEauthorblockN{
Ubaidullah Khan\IEEEauthorrefmark{1},
Raveen Wijewickrama\IEEEauthorrefmark{1},
Buddhi Ashan M.K.\IEEEauthorrefmark{1},
A. H. M. Nazmus Sakib\IEEEauthorrefmark{1},\\
Khoi Trinh\IEEEauthorrefmark{2},
Christina Duthie\IEEEauthorrefmark{1},
Nima Najafian\IEEEauthorrefmark{2},
Ahmer Patel\IEEEauthorrefmark{1},
R.N. Molina\IEEEauthorrefmark{1},\\
Anindya Maiti\IEEEauthorrefmark{2},
Sushil K. Prasad\IEEEauthorrefmark{1},
Greg P. Griffin\IEEEauthorrefmark{3},
Murtuza Jadliwala\IEEEauthorrefmark{1}
}
\IEEEauthorblockA{\IEEEauthorrefmark{1}University of Texas at San Antonio,
\IEEEauthorrefmark{2}University of Oklahoma,
\IEEEauthorrefmark{3}Oregon Department of Transportation}
\IEEEauthorblockA{Corresponding Author: Anindya Maiti (am@ou.edu)}
}

\begin{document}

\maketitle

\input{abstract}


\input{introduction}

\input{architecture}

\input{demo}

\bibliographystyle{plain}
\bibliography{references}

\end{document}

%% file: abstract.tex
\begin{abstract}
Micromobility vehicles, such as e-scooters, are increasingly popular in urban communities but present significant challenges in terms of road safety, user privacy, infrastructure planning, and civil engineering. Addressing these critical issues requires a large-scale and easily accessible research infrastructure to collect diverse mobility and contextual data from micromobility users in realistic settings. To this end, we present \name, a community research testbed comprising a fleet of customizable battery-powered micromobility vehicles retrofitted with advanced sensing, communication, and control capabilities. \name~enables interdisciplinary research at the intersection of computing, mobility, and urban planning by providing researchers with tools to design and deploy customized sensing experiments and access curated datasets. The testbed will enable advances in machine learning, privacy, and urban transportation research while promoting sustainable mobility.
\end{abstract}

%% file: introduction.tex
\section{Introduction}
\label{sec:intro}

Modern micromobility solutions, such as battery-powered single-rider bikes and e-scooters, have transformed short-distance commutes in urban communities \cite{scooter-pop1,scooter-pop2}. Due to their portable and user-friendly design, they do not require any operational training or license, and are easily available/accessible through popular rental service providers.
Its not surprising that they have emerged as a popular \emph{last-mile} transportation solution, especially in large and congested urban areas \cite{unagi-berkeley}. 
However, such a widespread adoption of battery-powered micromobility vehicles has also introduced significant challenges related to road safety, user privacy, sustainability, and urban infrastructure planning, requiring urgent attention from the research community.

Addressing these challenges requires a multidisciplinary approach, involving contributions from computer scientists, engineers (across computer, civil, environmental, and transportation disciplines), and urban planners. Although researchers have started responding to these emerging challenges with innovative solutions, the progress has been constrained by the scarcity of comprehensive, diverse, and open-access datasets on rider behavior and micromobility patterns. Such datasets, collected in real-world settings, are critical for developing robust solutions that can be effectively translated to urban contexts. 

Current micromobility research \cite{fang2018riders,orozco2022dockless,reck2021explaining} often relies on limited methodologies due to the challenges in accessing comprehensive and diverse datasets. Although partnering with commercial e-scooter providers (e.g., Bird \cite{bird_e_scooter} and Lime \cite{lime_e_scooter}) may seem like an ideal avenue for large-scale studies, such collaborations are rare. Commercial providers are often reluctant to share their data due to proprietary concerns, competitive interests, and privacy issues. Additionally, their datasets may not align with research objectives, as they typically focus on operational metrics, such as fleet performance and ridership statistics, rather than granular or exploratory data useful for research.
This lack of access forces researchers to resort to smaller-scale studies, often deploying limited numbers of participants and vehicles in controlled environments. While these studies provide some experimental control, they are inherently constrained by sample size, lack diversity, and require significant time and financial investment. The scarcity of open, high-quality experimental testbeds and datasets thus continues to impede the progress in understanding broader micromobility impacts and challenges.
Consequently, there is a growing need for a community-wide micromobility and urban data sensing testbed to advance research in this area.

\name~addresses these limitations by providing a highly-configurable micromobility testbed to conduct cost-effective experiments, featuring a large fleet of retrofitted e-scooters equipped with state-of-the-art sensing hardware. 
These vehicles, available to a large pool of participants across multiple campuses at the University of Texas at San Antonio (UTSA), enable research at scale and in realistic, diverse settings, presenting an invaluable resource for multidisciplinary research.

%% file: architecture.tex
\begin{figure*}[t] 
\centering
\includegraphics[width=0.95\linewidth]{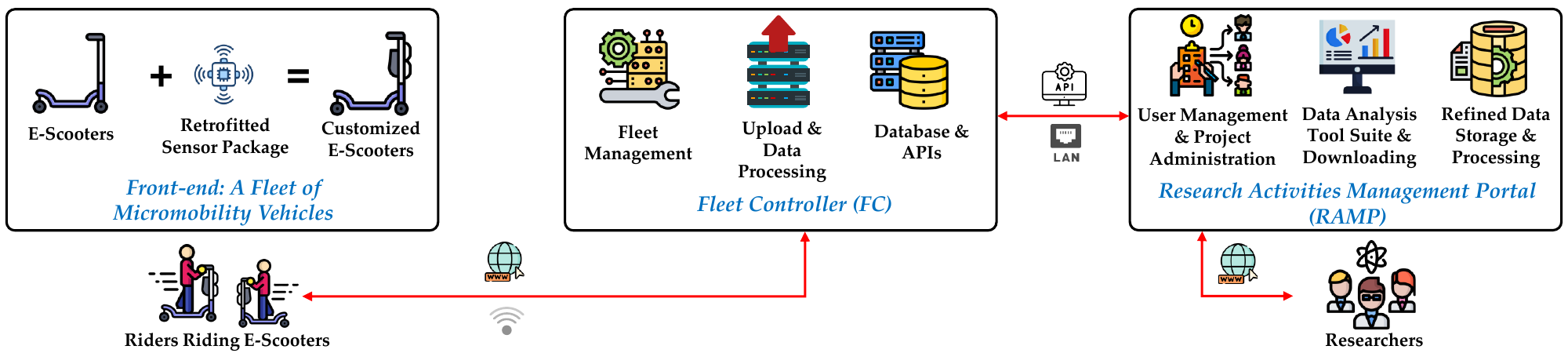} %
\caption{\name~architecture.}
\label{fig:overview}
\end{figure*}

\section{\name~Architecture}
\label{sec:archi}

At its core, \name~consists of a fleet of commercially available e-scooters enhanced with specialized sensors, which are provided to real-world participants for their micromobility journeys.
\name~ infrastructure (see \cref{fig:overview}) comprises three main components: the Wireless Base Station Computer (WBSC) integrated on e-scooters for sensing; the Fleet Controller (FC), which manages the fleet of e-scooters and stores raw data; and the Research Activities Management Portal (RAMP), which provides an interface for external researchers to engage with \name~and its data.
We define the following key terms in the ScooterLab system; \textit{project}: a data collection experiment,
\textit{researcher}: initiates experiments by requesting a project through RAMP, detailing the required data collection criteria, \textit{rider}: a volunteer operating an e-scooter, \textit{trip}: a single e-scooter journey undertaken by a rider.

\subsection{E-Scooters \& Wireless Base-station Computer (WBSC)}
\label{sec:scooters}

\begin{figure}[h]
\centering
\begin{subfigure}{0.35\linewidth}
    \centering
    \includegraphics[width=\linewidth]{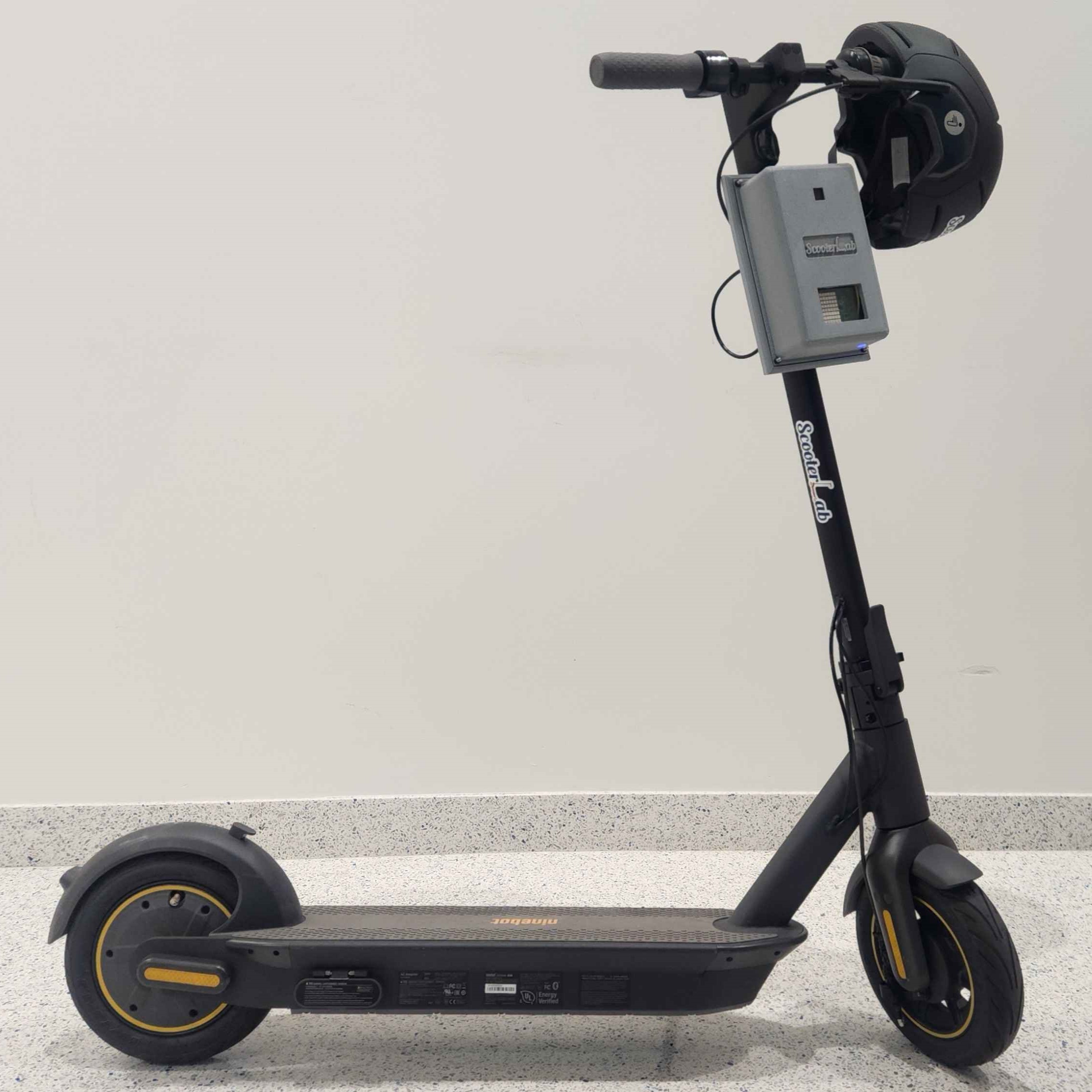}
    \caption{}
    \label{fig:scooter1}
\end{subfigure}
\hspace{0.05\linewidth}
\begin{subfigure}{0.35\linewidth}
    \centering
    \includegraphics[width=\linewidth]{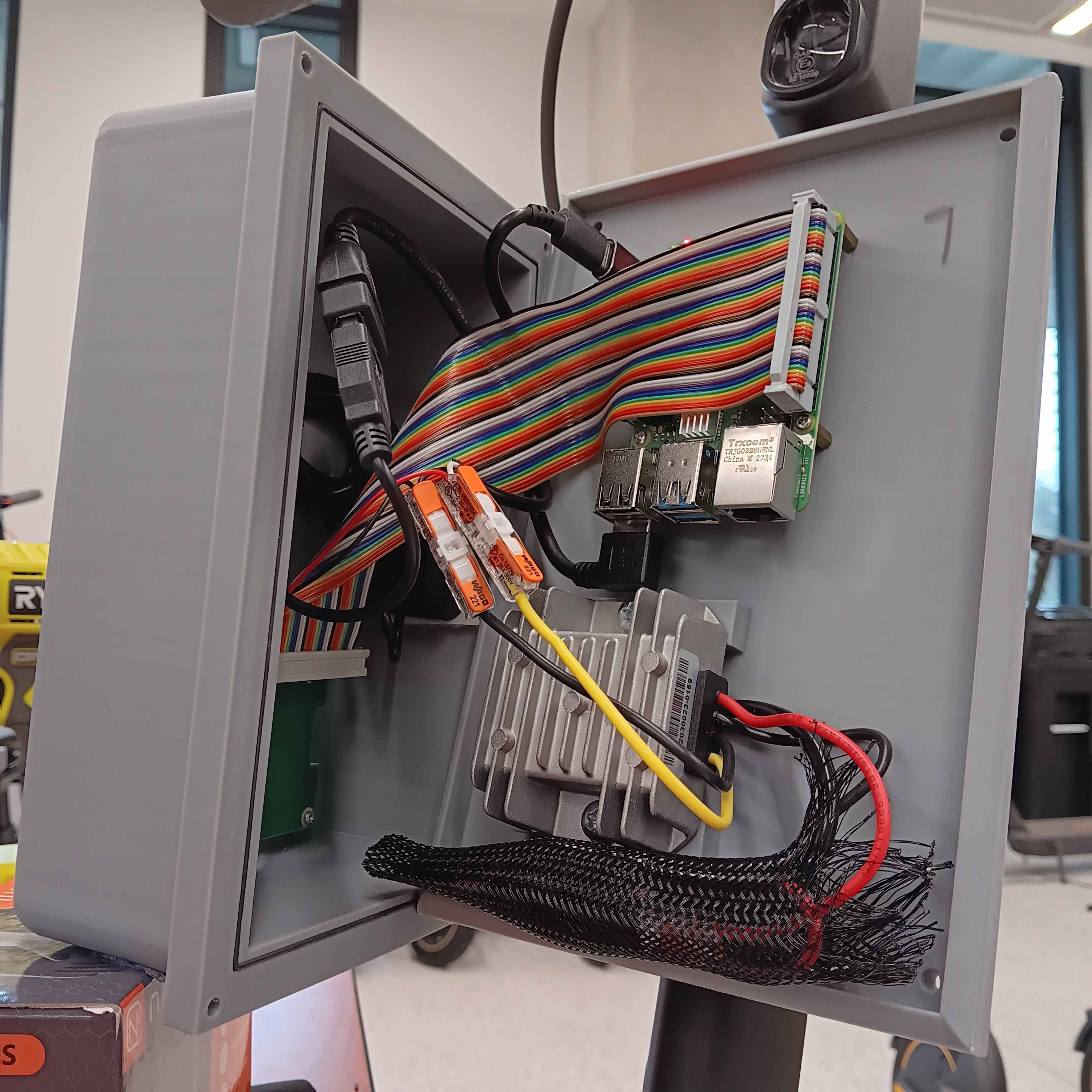}
    \caption{}
    \label{fig:scooter2}
\end{subfigure}
\caption{WBSC mounted \name~e-scooters.}
\label{fig:scooter}
\end{figure}

The WBSC serves as the computing unit for sensing and data transmission at the e-scooter level, enabling continuous data collection during rides (see \cref{fig:scooter}). It communicates with the FC to transmit recorded sensor data and receive configuration updates.

We use a Raspberry Pi 4 (4GB) as the computing unit of the WBSC due to its versatility in supporting multiple sensors through GPIO and USB interfaces.
A Sense HAT V2 is connected through the GPIO interface of the Pi to provide a base set of sensors, which features a gyroscope, an accelerometer, a magnetometer, a temperature sensor, a barometric pressure sensor, a humidity sensor, and a light sensor.
For location tracking, the WBSC integrates an Adafruit Ultimate GPS module via the USB interface, while a Pi Camera Module 3 and USB microphone enable audio and video sensing.
To meet evolving research or future \textit{project} needs, we designed the WBSC to support the addition of custom sensors via its GPIO and USB interfaces, enhancing the platform's adaptability to diverse experimental requirements.
Further, by powering the WBSC directly from the e-scooter's battery using a 40V to 5V DC/DC step-down converter, we eliminate the need for separate charging or external power, simplifying operations for \textit{riders}.
In terms of software, the Raspberry Pi operates on Debian 12 Pi OS, running a custom Python application that manages the WBSC's operations.
The WBSC temporarily stores trip data in local storage and automatically uploads it to the FC whenever the scooter comes within range of the UTSA Wi-Fi network.
The current \name~fleet consists of eight Segway G30 Max off-the-shelf e-scooters, which offer a range of up to 40 miles and a maximum speed of 18 mph. We plan to expand the \name~fleet to 80–100 e-scooters over the next two years.
Additionally, \name~offers a mobile application that allows riders to monitor scooter battery levels and range information.

\subsection{Fleet Controller (FC)}
\label{sec:fc}

The FC serves as the central hub for managing \name's e-scooter fleet, operational safety and data handling, thereby enabling seamless communication between e-scooters and the RAMP. 
FC offers the following three services.
The fleet management service enables the generation and deployment of configuration files for e-scooters, allowing dynamic adjustments to \textit{project} specific parameters such as modifying sampling rates or enabling specific sensors. 
The upload and data processing service handles the ingestion and preprocessing of \textit{trip} data received from the e-scooters.
Additionally, this service integrates external data sources, including traffic \cite{met_weather} and weather \cite{traffic_api} information. These datasets can potentially enhance the contextual understanding of mobility patterns by providing insights into external factors such as road conditions, weather-induced mobility variations, and traffic density. We will also accommodate the integration of additional external data sources to meet evolving research requirements.
Finally, the database and API service enables the storage of trip data and facilitates its retrieval for the RAMP.

The FC system runs on a server equipped with dual Intel Xeon 12-core processors, 512GB RAM, and 720TB storage.
A Flask-based application receives raw files from e-scooters, ensuring reliable and efficient data ingestion into the system, which is then stored in a MySQL database. A separate Flask-based API facilitates communication between the FC and RAMP, enabling real-time data transfer and experiment-specific queries. A 10 Gbps LAN interconnects the FC and RAMP systems, ensuring low latency and high throughput for API calls, enabling rapid access and processing of large datasets.

\subsection{Research Activities Management Portal (RAMP)}
\label{sec:ramp}
The RAMP is a software interface that serves research communities and \name~staff to interact with the testbed and the collected data.   
It primarily enables researchers to request new \textit{projects} and track their progress by providing tools to analyze, visualize, and download the collected data. 

We provide two data analysis tools for researchers to organize trip data via RAMP.
The \textit{Map} tool visualizes the collected data by rendering e-scooter trips that belong to a project along with the sensor data over a base map. The \textit{Stats} tool displays collected data using tabular and chart views. 
These tools provide data querying capabilities based on the spatiotemporal properties of \textit{trips} and their associated sensor data. Furthermore, they can export \textit{trip} data, enabling researchers to seamlessly integrate the data into their analysis workflows.

\begin{figure}[htp] 
\centering
\includegraphics[width=0.75\linewidth]{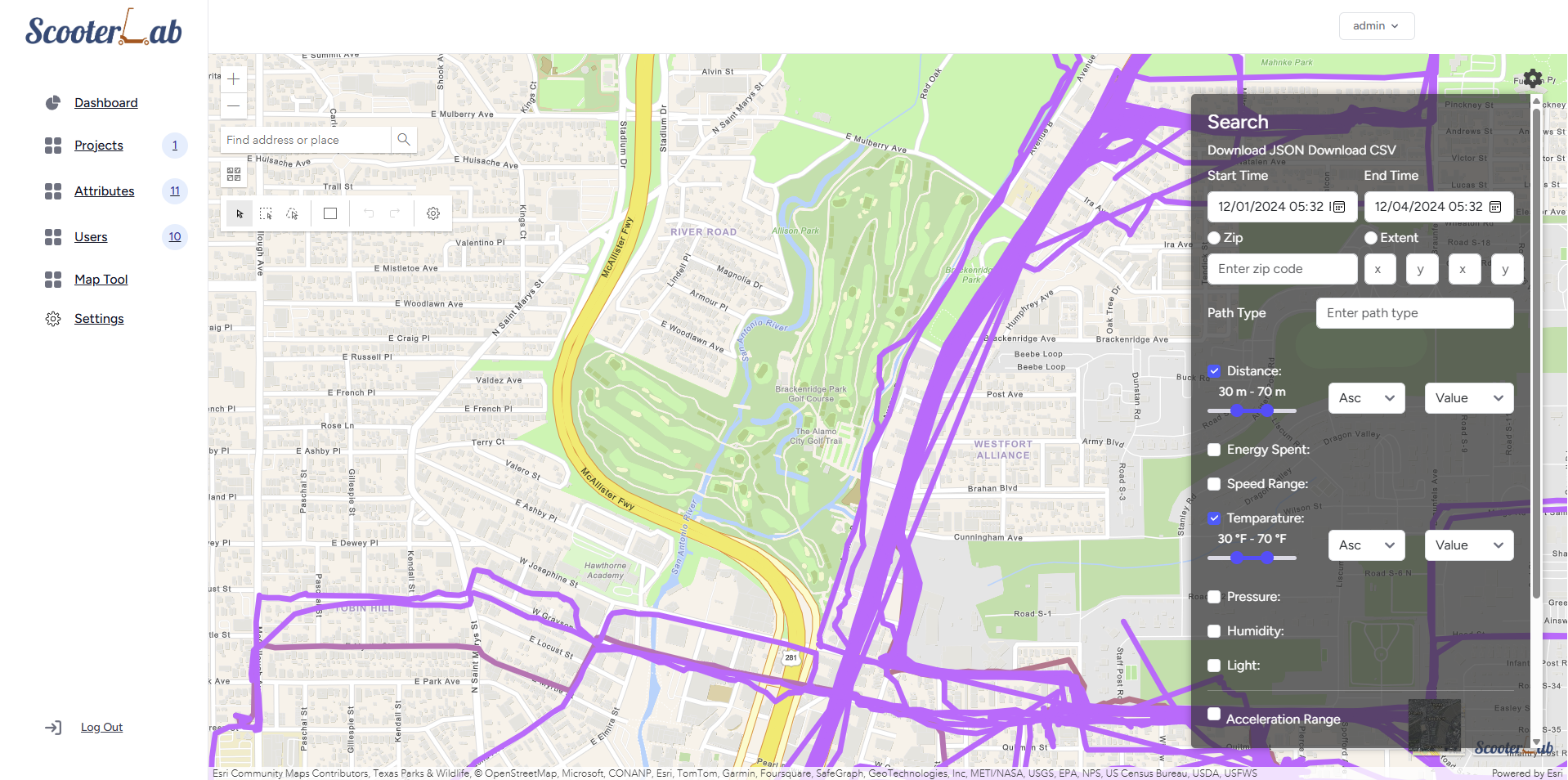}
\caption{RAMP \textit{Map} tool.}
\label{fig:ramp}
\end{figure}

RAMP system runs on a server equipped with two Intel Xeon Platinum 8468 processors, 2TB of memory, and four Nvidia L40S GPUs. It comprises a Laravel 10-based back-end, a MySQL database to manage user data, and JavaScript front-end interfaces. 
RAMP consists of the following three modules.
The user module provides services for authenticating, creating, and changing user information. 
The project module consists of functionalities to initiate new projects by allocating a fleet of e-scooters and defining the data collection policy. The data collection policy defines the list of sensors, the targeted geographic area, and the schedule for the project. 
Finally, the tool suite module includes the \textit{Map} tool, built using the ArcGIS Maps SDK for JavaScript, and the \textit{Stats} tool, which uses JavaScript to dynamically generate interactive tables and charts (see ~\cref{fig:ramp}).
In future, we plan to extend the RAMP tool suite to allow the researchers to perform computationally intensive analytics.

\subsection{Operationalization}
Individuals interested in using \name's e-scooters begin by scheduling a check-out appointment at a designated UTSA campus location, where they sign an informed consent form including \name's usage terms and conditions. They also view a mandatory video regarding safe e-scooter operation followed by completing a survey on their travel habits over the past seven days. 
Currently, participants can loan the e-scooters for up to two weeks, after which they can return them or renew them by going through the informed consent process again.
Prior to each loan or renewal, \name~staff inspect the e-scooters to ensure that they are in optimal working condition.

These operational and data collection procedures have been reviewed and approved by the Institutional Review Board (IRB) of UTSA. 
Future \textit{project} requests from \textit{researchers} may require modifications to these procedures or new IRB approvals to accommodate specific experimental designs or data collection requirements.

%% file: demo.tex
\section{Demonstration}
\label{sec:demo}

In this demonstration, we showcase the end-to-end capabilities of the ScooterLab platform, highlighting its potential as a micromobility sensing and research testbed. 
The demonstration begins with a walkthrough of the e-scooter hardware, highlighting the retrofitted sensors and the WBSC. 
We will also conduct an on-site data collection, highlighting e-scooter \textit{trip} data collection and its subsequent upload to the FC. 
Further, we will demonstrate how \textit{researchers} can leverage the \textit{Map} and \textit{Stats} tools offered by RAMP to create tailored datasets, query datasets specific to \textit{projects}, visualize e-scooter trips on a geospatial map, and perform basic data analysis tasks.
The ScooterLab platform will be accessible to researchers across disciplines, providing public access to curated datasets and tools to advance diverse studies in micromobility and related fields.